\documentclass[12pt]{article}
\usepackage{graphicx}
\usepackage{amssymb}
\usepackage{epstopdf}
\usepackage{multirow}
\newcommand{\raw}{\rightarrow}

\newcommand\mathC{\mkern1mu\raise2.2pt\hbox{$\scriptscriptstyle|$}
        {\mkern-7mu\rm C}}              
\newcommand{\mathR}{{\rm I\! R}}         

\newcommand{\be}{\begin{equation}}
\newcommand{\ee}{\end{equation}}

\let\ssection=\section
\renewcommand{\section}{\setcounter{equation}{0}\ssection}

\begin{document}
\begin{center}
{\large\bf Laws, Causation and Dynamics at Different Levels}
\end{center}

\begin{center}
Jeremy Butterfield \\
Trinity College, Cambridge CB2 1TQ: jb56@cam.ac.uk
\end{center}

\begin{center}  Published in {\em Interface Focus (Royal Society, London)}, {\bf 2}, 2012, pp. 101-114; (a special issue on top-down causation). doi: 10.1098/rsfs.2011.0052 
\end{center}

\begin{center}  6 August 2011 \end{center}

\begin{abstract}
I have two main aims. The first is general, and more philosophical (Section 2). The second 
is specific, and more closely related to physics (Sections 3 and 4).

The first aim is to state my general views about laws and causation at different `levels'. 
The main task is to understand how the higher levels sustain notions of law and causation  
that `ride free' of reductions to the lower level or levels. I endeavour to relate my views 
to those of other symposiasts. 

The second aim is to give a framework for describing dynamics at different levels, 
emphasising how the various levels' dynamics can mesh or fail to mesh. This framework is 
essentially that of elementary dynamical systems theory. The main idea will be, for 
simplicity, to work with just two levels, dubbed `micro' and `macro' which are related by 
coarse-graining. I use this framework to describe, in part, the first four of Ellis' five 
types of top-down causation.

\end{abstract}
\newpage
\tableofcontents

\newpage

\section{Introduction}\label{intro} 
I have two main aims. The first is general, and more philosophical  (Section 2). It 
concerns, not just this special issue's topic, top-down causation, but the general 
relations between `levels'. The second aim is specific, and more closely related both to 
top-down causation and to physics, in particular dynamical systems theory (Sections 3 and 
4).

In discussing relations between levels, I will take it that the overall task is to 
understand how the higher levels sustain notions of law, causation and explanation   that 
are `autonomous', or `ride free', from whatever reductions there might be to the lower 
level or levels; (or at least: notions that {\em seem} to be autonomous or to ride free!).  
This is a large task, with a large literature of controversy, both nowadays and in the 
past. The reasons for the controversy are obvious. People disagree about how to understand 
the notions of level and reduction, and also those of law, causation and explanation. They 
disagree about the extent to which, and the sense in which, the higher levels are 
autonomous or ride free. And these disagreements are  fuelled by having different sets of 
scientific examples in mind.

These disagreements become more vivid (and more comprehensible), when one considers 
historical changes in the disputants' scientific examples. One broad example is the demise 
of vitalism. Before the century-long rise of microbiology, biochemistry and molecular 
biology (from say 1860 to 1960), it was perfectly sensible to believe that biological 
processes depended on certain `vital forces'; and therefore that, as the slogan puts it, 
`biology is not reducible to chemistry and physics'---in a  much stronger sense of `not 
reducible' than one could believe today. Other broad examples extending over many decades 
are: (i) the rise of atomism and statistical mechanics in explaining irreversible 
macroscopic processes; and (ii) the rise of atomism, the periodic table and then quantum 
chemistry, in explaining chemical bonding. Again: before these developments, one could 
believe in irreducibility, e.g. of chemical bonding to physics, in a much stronger sense 
than one can today. In short: what historians now call `the second scientific revolution' 
from 1850 onwards has given us countless successful reductions of behaviour (both specific 
processes and general laws) at a higher (often macroscopic) level to facts at a lower 
(often microscopic) level.

Thus the overall philosophical task, both nowadays and in yesteryear, is: first, to state 
and defend notions of level and reduction, and of law, causation and explanation; and 
second, to use them to assess, in a wide range of contemporary scientific examples, the 
extent to which, and the sense in which, the higher levels are autonomous, or ride free, 
from the lower levels. But nowadays, after the triumphs of the second scientific 
revolution, we must expect the extent of, and-or senses for, such autonomy of the higher 
levels to be more restricted and-or more subtle. One aspect of this overall task is the 
topic of this special issue: assessing the prospects for top-down causation.

My own contribution will proceed in two stages. First (Section \ref{RSC}), I will summarize 
some of my own views about the overall task, relating them to top-down causation and the 
views of some other authors. For example, I will briefly endorse some views of Sober's 
about reduction and causation, and  List and Menzies' recent defence of top-down causation 
(Sections \ref{mra} and \ref{cause}). My overall views are defended in detail elsewhere 
(2011, 2011a). They mostly concern emergence, reduction and supervenience; (Section 
\ref{RSC} will report my construals of these contested terms).  I should  admit at the 
outset that I will have nothing distinctive to say about the notions of law, causation and 
explanation. But in fact, I take a broadly Humean view of all three; and it will be clear 
that this will fit well with my views on emergence, reduction and supervenience.

Second, in Sections \ref{mesh} and \ref{five}, I will give a framework for describing 
dynamics at different levels, emphasising how the various levels' dynamics can mesh, or 
fail to mesh. This framework is essentially that of elementary dynamical systems theory. 
The main idea will be, for simplicity, to work with just two levels, dubbed `micro' and 
`macro' which are related by coarse-graining. I then consider two topics, in Sections 
\ref{mesh} and \ref{five} respectively.

First, there is the question whether a micro-dynamics, together with a coarse-graining 
prescription, induces a well-defined macro-dynamics (Section \ref{mesh}). I describe how 
physics provides some precise and important examples of such `meshing' (e.g. in statistical 
mechanics), as well as examples where it fails. I will stress that failure of meshing need 
not be a problem, let alone a mystery: the pilot-wave theory, in the foundations of quantum 
mechanics, will provide a non-problematic example. I also discuss how to secure meshing by 
re-defining the coarse-graining; and relate the topic to the philosophical views of Fodor 
and Papineau on multiple realizability, and of List on free will.

Second, I use the framework to describe, in part, the first four of Ellis' (2008, 2012) 
five types of top-down causation (Section \ref{five}). There are various choices to be made 
in giving such a dynamical systems description of Ellis' typology; but I maintain that the 
fit is pretty good. In particular, I note that Ellis calls my first topic above, i.e. the 
meshing of micro- and macro-dynamics, `coherent higher level dynamics', or `the principle 
of equivalence of classes'; and he takes it as a presupposition of his typology of top-down 
causation.

Finally, a clarification. This paper has some ``reductionist'' features, which might be 
misleading. Thus in Section \ref{RSC} I will join Sober and Papineau in rejecting the 
multiple realizability argument against ``reductionism''. And in Section 4, I will not try 
to articulate the differences between my formal descriptions, in the jargon of dynamical 
systems, of Ellis' types of top-down causation, and Ellis' own informal and richer  
descriptions. These features might suggest that I deny any or all of the following three 
claims:---\\
\indent (i) There are, or can be, laws and-or explanation and-or causation at ``higher 
levels'', or in the special sciences. \\
\indent  (ii) There is a good notion of causation beyond that of functional dependence of 
one quantity on another.  \\
\indent (iii) Top-down causation, at least of Ellis' types, needs more than my dynamical 
systems framework.\\
But in fact, I endorse (i)-(iii). It is just that they are not centre-stage in my 
discussion.

\section{Reduction, supervenience and causation}\label{RSC}
My first aim is to summarize some of my views about the relations between levels. Sections 
\ref{redn} and \ref{mra} discuss reduction and `multiple realizability', and Section 
\ref{12C} discusses supervenience. Broadly speaking, I will deny the widespread (perhaps 
even orthodox?) views that multiple realizability prevents reduction, and that levels are 
typically related by supervenience without reduction. Section \ref{cause} concerns 
causation: here I will endorse Shapiro and Sober's, and List and Menzies', recent arguments 
for top-down causation.
 
\subsection{Reduction}\label{redn}
I have analyzed the relations between reduction, emergence and supervenience, elsewhere 
(2011, 2011a). In short, I construe these notions as follows. I take emergence as a 
system's having behaviour, i.e. properties and-or laws, that is novel and robust relative 
to some natural comparison class. Typically, the behaviour concerned is collective or 
macroscopic; and it is novel compared with the properties and laws that are manifest in the 
(theory of) the microscopic details of the system. I take reduction as a relation between 
theories: viz. deduction using appropriate auxiliary definitions. (As we will see, this is 
in effect a strengthening of the traditional Nagelian conception of reduction.) And I take 
supervenience as a weakening of this concept of reduction, viz. to allow infinitely long 
definitions; (more details in Section \ref{12C}). 

Then my main claim was that, with these meanings, emergence is logically independent both 
of reduction and of supervenience. In particular, one can have emergence with reduction, as 
well as without it. Physics provides many such examples, especially where one theory is 
obtained from another by taking a limit of some parameter. That is: there are many examples 
in which we deduce a novel and robust behaviour, by taking the limit of a 
parameter.\footnote{My (2011a) analysed four examples.  Footnote 6 and Section \ref{21B} 
will mention yet other examples.}  And emergence is also independent of supervenience: one 
can have emergence without supervenience, as well as with it.

Broadly speaking, this main claim gives some support to the ``autonomy'' of higher levels 
(cf. claim (i) at the end of Section \ref{intro}), viz. by reconciling such autonomy with 
the existence of reductions to lower levels. Some of my other claims had a similar 
reconciling intent: e.g. my joining Sober and Papineau in holding that multiple 
realizability is no problem for reductionism.\footnote{There were other claims I will not 
need here, e.g. that  emergence does not require limits, in particular not  ``singular'' 
limits.} I shall develop this position a little by discussing Nagelian reduction (this 
Subsection), multiple realizability (Section \ref{mra}) and supervenience  (Section 
\ref{12C}).

Nagel's idea is that reduction should be modelled on the logical idea of one theory being a 
{\em  definitional extension} of another.\footnote{The main source is Nagel (1961, pp. 
351-363); cf. also Hempel (1966, Chapter 8). Schaffner (2011) is a masterly review not only 
of Nagel's position, but also of others' critiques, defences and modifications of Nagel.}  
Writing $t$ for `top' and $b$ for `bottom', we say: $T_t$ is a
{\em definitional extension\/} of $T_b$, iff one can add to
$T_b$ a set $D$ of definitions, one for each of $T_t$'s non-logical symbols,
in such a way that $T_t$ becomes a sub-theory of the augmented theory $T_b \cup D$.
That is: In the augmented theory, we
can prove every theorem of $T_t$. Here, a {\em definition} is a statement, for a predicate, 
of co-extension; and for a singular term, of co-reference. To be precise: for a predicate 
$P$ of $T_t$ it would be a universally quantified biconditional with $P$ on the left hand 
side stating that $P$ is co-extensive with a right hand side that is a open sentence $\phi$ 
of $T_b$ built using such operations as Boolean connectives and quantifiers. Thus if $P$ is 
$n$-place:  $(\forall x_1)...(\forall x_n)(P(x_1,...,x_n) \equiv \phi(x_1,...,x_n))$. (The 
definitions are often called `bridge laws', or `bridge principles'.)

A {\em caveat}. I said that Nagel held that reduction `should be modelled on' the idea of 
definitional extension', because definitional extension is (a) sometimes too weak as a 
notion of reduction, and (b) sometimes too strong. 

As to (a): Nagel  (1961, pp. 358-363) holds that the reducing
theory $T_b$ should explain the reduced theory $T_t$; and following
Hempel, he conceives explanation in deductive-nomological
terms. Thus he says, in effect, that $T_b$ reduces $T_t$
iff:\\
\indent  (i): $T_t$ is a definitional extension of $T_b$; and \\
\indent (ii): In each of the definitions of $T_t$'s terms, the {\em definiens} in the 
language of $T_b$ must play
a role in $T_b$; so it cannot be, for example, a
heterogeneous disjunction. 

As to (b): Definitional extension is sometimes too  strong as a notion of reduction; as 
when  $T_b$ {\em corrects}, rather than implies, $T_t$. Thus Nagel says that a case
in which $T_t$'s laws are a close approximation to
what strictly follows from $T_b$ should count as reduction, and be called `approximative 
reduction'.

More important for us is the fact that definitional extensions, and thereby Nagelian 
reduction, can perfectly well accommodate what philosophers call {\em functional 
definitions}. These are  definitions of a predicate or other non-logical symbol (or in 
ontic, rather than linguistic, jargon: of a property, relation etc.) that are second-order, 
i.e. that quantify over a given `bottom set' of properties and relations. The idea is that 
the {\em definiens} states a pattern among such properties, typically a pattern of causal 
and lawlike relations between properties. So an $n$-tuple of bottom properties that 
instantiates the pattern in one case is called a  {\em realizer} or {\em realization} of 
the {\em definiendum}. And the fact that in different cases, different such $n$-tuples 
instantiate the pattern is called {\em multiple realizability}. Examples of functional or 
second-order properties, and so of multiple realizability, are legion. For example: the 
property of being locked is instantiated very differently in padlocks using keys, 
combination locks etc.

\subsection{The multiple realizability argument refuted}\label{mra}
Multiple realizability is undoubtedly a key idea, philosophically and scientifically, for 
our overall task: understanding relations between levels, and especially how higher levels 
can be autonomous, or ride free, from lower levels. Agreed, there is not much to be said by 
way of a theory about being locked: and similarly for countless other multiply realizable 
properties, like being striped, or being mobile, or being at least 50 per cent metallic or 
...  For being locked, being striped etc. do not define, or contribute to defining, 
significant levels. But some multiply realizable properties do so: cf. claim (i) at the end 
of Section \ref{intro}.

Here is a schematic biological example; (my thanks to a referee). Fitness is multiply 
realized by the different morphological, physiological and behavioral properties of 
organisms. Indeed, {\em very} multiply realized: what makes a cockroach fit is very 
different from what makes a daffodil fit. Thus fitness is a higher-order or `more abstract' 
similarity of organisms; (as are its various degrees). And unlike being locked etc., it 
contributes to defining a significant level: there are general truths about it and related 
notions, to be expressed and explained. For this it is not enough to have a theory about 
cockroaches, and another one for daffodils etc. Rather, we need the theory of natural
selection.

In short: multiple realizability is undoubtedly important for understanding relations 
between levels. But many philosophers go further than this. Some think that multiple 
realizability provides an argument against reduction. The leading idea is that the {\em 
definiens} of a multiply realizable property shows it to be too ``disjunctive'' to be 
suitable for scientific explanation, or to enter into laws. And some philosophers think 
that multiple realizability prompts a non-Nagelian account of reduction; even suggesting 
that definitional extensions cannot incorporate functional definitions. 

I reject both these lines of thought. Multiple realizability gives no argument against 
definitional extension; nor even against stronger notions of reduction like Nagel's, that 
add further constraints additional to deducibility, e.g. about explanation. That is: I 
believe that such constraints are entirely compatible with multiple realizability. This was 
shown very persuasively by Sober (1999). But since these errors are unfortunately 
widespread, it is worth first rehearsing, then refuting, the multiple realizability 
argument.

We again envisage two theories $T_b$ and $T_t$, or two sets of properties, $\cal B$ and 
$\cal T$, defined on a set $O$ of objects. The choice between theories and sets of 
properties makes almost no difference to the discussion; and I shall here mostly refer to 
$\cal B$ and $\cal T$, rather than $T_b$ and $T_t$. So multiple realizability means that  
the instances in $O$ of some `top' property $P \in {\cal T}$ are very varied 
(heterogeneous) as regards (how they are classified by) their properties in $\cal B$.

The {\em multiple realizability argument} holds that in some cases, the instances of $P$ 
are so varied that even if there is an extensionally correct definition of $P$ in terms of 
$\cal B$, it will be so long and-or heterogeneous that:\\
\indent \indent (a): explanations of singular propositions about an instance of $P$ cannot 
be given in terms of $\cal B$, whatever the details about the laws and singular 
propositions involving $\cal B$;\\
\indent \indent \indent \indent and-or; \\
\indent \indent (b): $P$ cannot be a natural kind, and-or cannot be a law-like or 
projectible property, and-or cannot enter into a law,  from the perspective of $\cal B$.

\indent Usually an advocate of (a) or (b) is not `eliminativist', but rather 
`anti-reductionist'. $P$ and the other properties in $\cal T$ satisfying (a) and-or (b) are 
not to be eliminated as cognitively useless. Rather, we should accept the taxonomy they 
represent, and thereby the legitimacy of explanations and laws invoking them.  Probably the 
most influential advocates have been: Putnam (1975) for version (a), with the vivid example 
of a square peg fitting a square hole, but not a circular one; and Fodor (1974) for version 
(b), with the vivid example of $P$ = being money.

I believe that Sober (1999) has definitively refuted this argument, in its various 
versions, whether based on (a) or on (b), and without needing to make contentious 
assumptions about topics like explanation, natural kind and law of nature. As he shows, it 
is instead the various versions of the argument that make contentious assumptions! I will 
not go into details. Suffice it to make three points, by way of summarizing Sober's 
refutation.\footnote{Agreed, in philosophy, there is always more to say. I do not pretend 
that Sober's paper is the last word on the subject: in a large literature, I recommend 
Shapiro (2000, especially Section IVf. p. 643f.) and Papineau (2010, especially Section 4, 
pp. 183f.).} The first two correspond to rebutting (a) and (b); the third point is broader 
and arises from the second.

\indent As to (a): the anti-reductionist's favoured explanations in terms of $\cal T$ do 
{\em not} preclude the truth and importance of explanations in terms of $\cal B$. As to 
(b): a disjunctive definition of $P$, and other such disjunctive definitions of properties 
in $\cal T$, is no bar to a deduction of a law, governing $P$ and other such properties in 
$\cal T$, from a theory $T_b$ about the properties in $\cal B$. Nor is it a bar to this 
deduction being an explanation of the law. 

The last sentence of this refutation of (b) returns us to the question whether to require 
reduction to obey further constraints apart from deduction. The tradition, in particular 
Nagel himself, answers Yes; (as I reported in {\em caveat} (a), Section \ref{redn}). Nagel 
in effect required that the {\em definiens}  play a role in the reducing theory $T_b$. In 
particular, it cannot be a very heterogeneous disjunction. (Recall: the {\em definiens} is 
the right-hand-side of a bridge principle.) The final sentence of the last paragraph 
conflicts with this view. At least, it conflicts if this view motivates the 
non-disjunctiveness requirement by saying that non-disjunctiveness is needed if the 
reducing theory $T_b$ is to explain the laws of reduced theory $T_t$. But I reply: so much 
the worse for the view. Sober puts this reply as a rhetorical question (1999, p. 552): `Are 
we really prepared to say that the truth and lawfulness of the higher-level generalization 
is {\em inexplicable}, just because the ... derivation is peppered with the word `or'?' I 
agree with him: of course not!

\subsection{Supervenience? The need for precision}\label{12C}
So far my main points have been that reduction in a strong Nagelian sense is compatible 
with both emergence (Section \ref{redn}) and multiple realizability (Section \ref{mra}). 
But these points leave open the questions how widespread is reduction, and what are the 
other, perhaps typical or even widespread, relations between theories at different levels. 
My view is that, within physics and even between physics and other sciences, reduction---at 
least in the approximative sense mentioned in {\em caveat} (b) of Section \ref{redn}---is 
indeed widespread.  I develop this view in (2011, Section 3.1.2), (2011a, Sections 4f.), 
partly in terms of the unity of nature; (cf. Section \ref{intro}'s opening remarks about 
the second scientific revolution).

As to what relation or relations hold when reduction fails, philosophers' main suggestion 
has been: {\em supervenience}. Roughly speaking, this notion is a strengthening of the idea 
of multiple realizability. I will first explain the notion, and then state four misgivings 
about it.

Again we envisage two sets of properties, $\cal B$ and $\cal T$, defined on a set $O$ of 
objects. We say that $\cal T$ {\em supervenes on}  $\cal B$ (also: {\em is determined by} 
or {\em is implicitly defined by}  $\cal B$) iff any two objects in $O$ that
match in all properties in $\cal B$ also match in all
properties in $\cal T$. Or equivalently: any two objects that differ in a property
in $\cal T$ must also differ in some property or other in
$\cal B$. (One also says that $\cal B$ {\em subvenes}  $\cal T$.) A standard (i.e. largely 
uncontentious!) example takes $O$ to be the set of pictures, $\cal T$ their aesthetic 
properties (e.g. `is well-composed'), and $\cal B$ their physical properties (e.g. `has 
magenta in top-left corner').

It turns out that supervenience is a weakening of Section \ref{redn}'s notion of 
definitional extension; namely to allow that a definition in the set $D$ might have an {\em 
infinitely} long {\em definiens} using $\cal B$. The idea is that for a property $P \in 
{\cal T}$, there might be infinitely many different ways, as described using $\cal B$, that 
an object can instantiate $P$: but provided that for any instance of $P$, all objects that 
match it in their $\cal B$-properties are themselves instances of $P$, then supervenience 
will hold.

Thus many philosophers have held that in cases where one level or theory  seems irreducible 
to another, yet to be in some sense `grounded' or `underpinned' by it, the relation is in 
fact one of supervenience. They say that the irreducible yet grounded level or theory 
(specified by its taxonomy of properties $\cal T$) supervenes on the other one. That is, 
there is supervenience without definitional extension: at least one definition in $D$ is 
infinitely long.

At first sight, this looks plausible: recall from the start of Section \ref{mra} that 
examples of multiple realizability are legion. But we should note four misgivings about it. 
The first two are widespread in the literature; the third and fourth are more my own. The 
first and fourth are philosophical limitations of supervenience; the second and third, 
scientific limitations. 

First: philosophers of a  metaphysical bent who discuss reduction, emergence and related 
topics find it natural to require that in reduction, the `top' properties $\cal T$ are 
shown to be identical to properties  in (or perhaps composed from) $\cal B$; and that this 
is so, whether the reduction is finite, as in definitional extension, or infinite as in 
supervenience. But the identity of properties (and the principles for composing properties) 
are controversial issues in metaphysics; and the holding of a supervenience relation is not 
generally agreed to imply identity. So for such philosophers, supervenience leaves a major 
question unanswered.

Second: although the distinction between finite and infinitely long definitions is 
attractively precise, it seems less relevant to the issue whether there is a reduction than 
another, albeit vague, distinction: viz. the  distinction between definitions and 
deductions that are short enough to be comprehensible, and those that are not. Recall that 
according to Section \ref{redn}'s notion of a definitional extension, a definition in $D$ 
can be so long as to be incomprehensible, e.g. a million pages---to say nothing of the 
length of the deductions! 

Thus the remarks usually urged to show a supervenience relation in some example, e.g. that 
no-one knows how to construct a finite definition of  `is well-composed' out of `has 
magenta in top-left corner' and its ilk, are not compelling. Our inability to complete, or 
even begin, such a definition is no more evidence that  a satisfactory definition would be 
infinite, than that it would be incomprehensibly long. In other words: we have no reason to 
deny that the example supports a definitional extension, albeit an incomprehensibly long 
one. And so far as science is concerned, definitional extension with incomprehensibly long 
definitions and deductions is useless: that is, it may as well count as a failure of 
reduction. Philosophers, including Nagel himself, have long recognized this point: recall 
the {\em caveat} (a) in Section \ref{redn}.\footnote{
The idea of incomprehensible definitions and deductions, and thereby the need for 
higher-level concepts and laws, is often illustrated with cellular automata such as 
Conway's game of Life, with e.g. `glider' as a higher-level concept (cf. Dennett  (1991, 
pp. 196-200), Bedau (2003, pp. 164-178), O'Connor (2012, pp. ?1-3;). The idea is: these 
concepts and laws are  definable and deducible from Life's basic rules, but only by 
processes so grotesquely long that you would be ill-advised---mad!---to try and follow 
them, rather than investigating the higher-level behaviour directly.

Besides: a theorem in logic (Beth's theorem) shows that under certain conditions (viz. 
first-order finitary languages), the  finite-infinite distinction collapses in the sense 
that if every term of $T_t$ is implicitly definable in $T_b$, then $T_t$ is a definitional 
extension (i.e. with finite definitions) of $T_b$. This point was first emphasized by 
Hellman and Thompson; more details are in my (2011, Section 5.1).}

The third misgiving is similar to the second, in that both accuse supervenience of 
having---for all its popularity in philosophy---limited scientific value. But where the 
second sees supervenience's allowance of infinite disjunctions as a distraction from the 
more important issue of comprehensibility, the third sees supervenience's allowance of 
infinite disjunctions as a distraction from the more important issue of the limiting 
processes that occur in the mathematical sciences, and in particular in examples of 
emergence in physics. That is: because supervenience's infinity of `ways (in terms of $\cal 
B$) to be $P \in {\cal T}$' bears no relation to the taking of a limit (e.g. through a 
sequence of states, or of quantities, or of values of a parameter), it sheds little or no 
light on such limits, in particular on the emergent behaviour that they can produce. 
Agreed, this sort of accusation can only be made to stick by analyzing examples: suffice it 
to say here that my (2011a, Section 4f.) analyzes four such. 

The fourth misgiving concerns philosophers' appeal to supervenience, not as a relation 
between two independently specified levels or theories, but as a tool for precisely 
formulating physicalism: the doctrine that, roughly speaking, all facts supervene on the 
physical facts. Here, my complaint is: for physicalism to be precise,  you  need to state 
precisely what are `the physical facts' (or what is `the physical supervenience basis'). 
Sad to say: in the philosophical literature, both proponents and opponents of physicalism 
tend to be vague about this.  Here is one example which has been discussed widely; (more 
details in (2011, Section 5.2.2)). (I also recommend Sober's very original discussion of 
how the definition of, and our reasons for, physicalism are usefully cast in terms of 
probability, especially the Akaike framework for statistical inference (1999a, pp. 136-138, 
159-161, 163-168).)

If there had been fundamental many-body forces (called by C. D. Broad, the British 
emergentist of the early twentieth century: `configurational forces'), then the theory of a 
many-body system would not be supervenient on (let alone a definitional extension of) a 
theory of its components that used only two-body interactions.\footnote{Here I endorse 
Scerri's (2011, pp. ?3-4) critique of Hendry's surely maverick suggestion (2010, Section 3) 
that configurational forces {\em are} needed in modern quantum chemistry to explain 
molecular chirality and shape. On the contrary, I take it to be well established that the 
explanation lies in superselection (classical quantities) being rigorously emergent in 
appropriate limits (e.g. Primas (1981, pp. 335-342), Amann (1993, Section 5), Bishop and 
Atmanspacher (2006, Section 4.1). So this is another example of Section \ref{redn}'s 
reconciling claim that emergence and reduction are compatible.} Of course, if there had 
been such forces, physicists would have made it their business to investigate them, so that 
a phrase like `the physical facts' would have come to include facts about such 
configurational forces, as well as facts about the familiar two-body forces. At least it 
would have come to include such facts if the configurational forces turned out to fit into 
the familiar general frameworks of (classical or quantum) mechanics, e.g. having a precise 
quantitative expression as a term in a Hamiltonian. But this says more about the elasticity 
of the word `physics', or about universities' departmental structure, than about the truth 
of a substantive doctrine of `physicalism'!

\subsection{Causation}\label{cause}
So far I have ignored issues about time-evolution, and in particular causation: I have 
stressed what one might call `synchronic issues', rather than `diachronic issues'. But from 
now on, diachronic issues will be centre-stage. As I mentioned in Section \ref{intro}, I 
take a broadly Humean view of causation, but do not advocate a specific account. Nor will I 
need such an account for the rest of this paper's aims. There are three such aims. In this 
Subsection, I will report and recommend two recent arguments broadly in favour of top-down 
causation. My final aim, in later Sections, will take longer: it is to describe Ellis' 
types of top-down causation, in terms of functional dependence. (Cf. claims (ii) and (iii) 
at the end of Section \ref{intro}.)

Of course, there is much to say about top-down causation apart from what follows in the 
rest of this paper; and even apart from my fellow symposiasts---in a large literature, I 
recommend Bedau (2003, pp. 157-160, 175-178). And I cannot trace the consequences of what 
follows, for other authors' views. But I commend what follows to advocates of top-down 
causation, such as Ellis and Atmanspacher, Auletta, Bishop, Jaeger and O'Connor. For I 
think it makes precise some of their claims: such as that higher-level facts or events 
constrain, modify or form a context for the lower level, which is therefore not independent 
of the higher level (Ellis, Auletta, and Jaeger); or that lower-level facts or events are 
necessary but not sufficient for higher-level ones (Atmanspacher and Bishop).

So I turn to reporting and recommending the two arguments. The first is Shapiro and Sober's 
argument, not so much {\em for} top-down causation, as {\em against} a contrary position, 
viz. epiphenomenalism. This is the doctrine that higher-level states cannot be causes, i.e. 
they are causally ineffective. Thus the idea of epiphenomenalism is that such states are 
pre-empted, as causes, by lower-level states. (Here we could say `property', `fact' or 
`event', instead of `state': it would make no difference to what follows.) Shapiro and 
Sober rebut this, by adopting an account of causation in terms of intervention (or, in 
another jargon: in terms of manipulation). On the other hand, the second argument is List 
and Menzies' positive argument for top-down causation; it is based on an account of 
causation in terms of counterfactuals. Fortunately, both accounts of causation are 
plausible, and I will not need to choose between them. Nor will I need to develop the 
accounts' details. For the argument by each pair of authors needs only the basic ideas of 
the account.
 
I can present both arguments in terms of the same example of a pair of levels: the mental 
and the physical. Both arguments are very general, and apply equally to other examples of 
pairs of levels. But this example has various advantages. It is vivid and widely discussed 
in philosophy. All these four authors use it. Although Shapiro and Sober also discuss 
higher-level causation in biology, especially evolutionary biology, for List and Menzies, 
this example is the main focus. So their central case of top-down causation is mental 
causation, e.g. my deciding to raise my arm causing it to go up. Besides, all these authors 
rebut various formulations and defences of epiphenomenalism about the mental with respect 
to the physical by Kim, who is probably the most prolific recent writer on the 
mental-physical relationship.

Thus here, in terms of the mental and the physical, is what Shapiro and Sober call `the 
master argument for epiphenomenalism':
\begin{quote}     
How could believing or wanting or feeling  cause behavior?  Given that any instance of a 
mental property $M$ has a physical micro-supervenience base $P$, it would appear that $M$ 
has no causal powers in addition to those that $P$ already possesses.  The absence of these 
additional causal powers is then taken to show that the mental property $M$ is causally 
inert. (2007, p. 241; notation changed)
\end{quote}
In addition to Kim's formulations of this argument that Shapiro and Sober go on to 
document, compare Kim (1999, p. 149). The argument is a cousin of what is often called `the 
exclusion argument', also often advocated by Kim: for a discussion, cf. e.g. Humphreys 
(1997).

Assessing this argument depends of course on one's account of causation. (And, one might 
guess: on one's account of realization or supervenience---but in fact, the varieties of 
these notions turn out not to matter.) But the argument fails utterly, on each of two 
plausible accounts of causation: the interventionist account adopted by Shapiro and Sober, 
and the counterfactual account adopted by List and Menzies. And the failure follows from 
just the basic features of each pair of authors' account. 

Thus the leading idea of Shapiro and Sober's rebuttal is:
\begin{quote} 
To find out whether $M$ causally contributes to $N$, you manipulate the state of $M$ while 
holding fixed the state of any common cause $C$ that affects both $M$ and $N$; you then see 
whether a change in the state of $N$ occurs. ... {\em  It is not relevant, or even 
coherent, to ask what will happen if one wiggles $M$ while holding fixed the 
micro-supervenience base $P$ of $M$.} ... Because a supervenience base for $M$ provides a 
sufficient condition for $M$, where the entailment has at least the force of nomological 
necessity, asking this question leads one to attempt to ponder the imponderable---would $N$ 
occur if a sufficient condition for $M$ occurred but $M$ did not? (2007, pp. 238-240; 
notation changed) 
\end{quote}   
Besides, Shapiro and Sober strengthen this rebuttal by augmenting their account of 
causation with probabilities; (cf. also Sober 1999a, pp. 145-149). There is a happy 
concordance here between interventionist and probabilistic accounts of causation. But 
again, I will not need to give details.

I turn to List and Menzies (2009, 2010). In short, they make two points: (i) a general 
point, in common with other authors, which supports the idea of causation occurring at 
higher levels, and between levels (cf. claim (i) at the end of Section \ref{intro}); and 
(ii) the specific argument in favour of top-down causation, and against epiphenomenalism. 
Broadly speaking, their first point supports the idea that we should understand causation 
in terms of counterfactuals (especially `if $C$ had not occurred, then $E$ would not have 
occurred'). On the other hand, their specific argument assumes some such counterfactual 
account.\footnote{But again: that is a small price to pay, since such accounts are 
plausible, especially when compared with the traditional rival idea (called `nomological 
sufficiency') that a cause, taken together with the laws, should imply the effect.}  

The general point is that a cause needs to be specific enough to produce its effect---but 
not more specific. The point is clear from how we think about causes in countless examples. 
What caused the bull to be in a rage? Answer: the bull's seeing the matador's red cape 
nearby. If  the cape  happens to be crimson and the matador to be standing three metres 
from the bull, nevertheless the cause is as stated. It is not the more specific fact of the 
bull's seeing the matador's crimson cape three metres away. 

Nor is this point just a matter of our intuitive verdicts in countless examples. It is 
upheld by plausible accounts of causation, in terms of counterfactuals (stemming from Lewis 
1973). The key idea is that $C$'s causing $E$ requires that if $C$ were not to hold, then 
$E$ would not hold. And indeed: if the cape were not red, but say green, then the bull 
would not be enraged. But if $C$ is too specific, this requirement tends to fail. We cannot 
conclude that if the cape were not crimson, the bull would not be enraged---for if the cape 
were not crimson, it might well have been some other shade of red, and then the bull would 
still have been enraged.       

 This point applies in countless examples where the contrast between appropriate and 
too-specific causes corresponds to a contrast between levels; such as the mental and the 
physical. Witness the bull example (which is adapted from Yablo 1992); or in the other 
`direction', a mental state causing a physical one, as in the time-honoured example of 
arm-raising. What caused my arm to go up? My deciding to raise it.

To sum up: this point teaches us that more specific information about the facts and events 
in some example is {\em not} always illuminating about causal relations; and ({\em a 
fortiori}), that it is wrong to think that the `best' or `real' causal explanation of the 
facts and events always lies in the most specific and detailed information about them. (Of 
course, it is `reductionists' rather than others who are most likely to suffer these bad 
temptations.) 
And so this point supports the idea that there are causal relations between facts or events 
at higher levels, or between different levels.\footnote{Agreed, it does not by itself imply 
such relations. But nor should we expect  a general principle about causation to dictate on 
such specific matters as between which levels there are causal relations.}

List and Menzies build on this general point so as to refute the `master argument', that a 
mental state $M$ cannot cause a later physical, say neural, state $N$, since its realizing 
or subvening neural state $P$ pre-empts it as a cause. They show that on plausible accounts 
of causation invoking counterfactuals, the argument fails. Indeed, they state a causal 
assumption about how $M$ causes $N$, that in many actual cases we have good reason to 
believe true, and that implies that $P$ is {\em not} a cause of $N$. (This assumption is 
that even if $M$ were realized by a physical state other than its actual realizer $P$, $N$ 
could still obtain. List and Menzies' jargon is that $M$'s causing $N$ is 
`realization-insensitive'; cf. 2009, Section V.) 

To sum up: here is excellent news for advocates of top-down causation, at least if they 
like interventionist or counterfactual accounts of causation. More power to these authors' 
elbows. Or, expressed in the spirit of their results: more power to their wills, so as to 
cause their elbows to move!

\section{Dynamics at different levels}\label{mesh}
I now turn to this paper's second main aim. In this Section, I give a framework for 
describing dynamics at different levels, emphasising how two levels' dynamics can mesh or 
fail to mesh. Section \ref{five} then applies the framework to describe some of Ellis' 
(2008) types of top-down causation. 

There will of course be some connections with topics addressed in Section \ref{RSC}. Here 
are two examples. (1): Section \ref{21B} describes how Papineau's critique of Fodor's 
version of the multiple realizability argument (cf. Section \ref{mra}) is a matter of two 
levels' dynamics failing to mesh. (2): I admit at the start of Section \ref{five} that my 
descriptions of Ellis' types are partial, because they avoid controversies about what is 
required for causation, beyond functional dependence of quantities; (cf. claims (ii) and 
(iii) at the end of Section \ref{intro}).

\subsection{The framework introduced}\label{21A}
For simplicity, I will work with just two levels, dubbed `micro' and `macro', which are 
related by coarse-graining. There will be several other simplifying assumptions, as 
follows.\\
\indent (i): We think of the micro-level as a state space ${\mathbb S}$, with the 
micro-dynamics as a map $T:{\mathbb S} \raw {\mathbb S}$ (so time is discrete). Since $T$ 
is a function, we assume a past-to-future micro-determinism.\\
\indent (ii): But $T$ need not be invertible, so future-to-past determinism can fail. 
Besides, most of what follows applies if $T$ is just a binary relation, i.e. one-many as 
well as many-one, so that there is past-to-future micro-{\em in}determinism. (This 
indeterminism need not reflect quantum mechanics---under the orthodox interpretation! It 
could reflect the system being {\em open}.) \\
\indent (iii): We will not need to discuss in detail the set $\mathbb Q$ of quantities. But 
we envisage that each $Q \in {\mathbb Q}$ is a $\mathR$-valued function on $\mathbb S$. For 
a classical physical state $s$, $Q(s)$ would be thought of as the system's intrinsic or 
possessed value for $Q$ when in $s$. But in quantum theory, $Q(s)$ would naturally be taken 
to be a Born-rule expectation value, e.g. $s$ is a Hilbert space vector, and $Q(s) := 
\langle s | {\hat Q} | s \rangle$. In either classical or quantum physics, we naturally 
think of $\mathbb Q$ as separating $\mathbb S$ in the sense that for any distinct $s_1 \neq 
s_2 \in {\mathbb S}$, there is a $Q \in {\mathbb Q}$ such that $Q(s_1) \neq Q(s_2)$.\\
\indent (iv): We think of the macro-level as given by a partition $P$ of $\mathbb S$, i.e. 
a decomposition of $\mathbb S$ into mutually exclusive and jointly exhaustive subsets $C_i 
\subset {\mathbb S}$. The $C_i$, i.e. the cells of the partition $P = \{ C_i \}$, are the 
{\em macro-states}. So $C$ stands for `{\em cell}' or `{\em coarse-graining}'.\\
\indent (v): Filling out (ii)-(iv): we naturally think of each $C_i$ as specified by a set 
of values for some subset of $\mathbb Q$, the macro-quantities ${\mathbb Q}_{\rm{mac}} 
\subset {\mathbb Q}$. That is: each $C_i$ is the intersection of level-surfaces, one for 
each quantity in ${\mathbb Q}_{\rm{mac}}$. But we can downplay quantities, and just focus 
on the macro-states $C_i$, i.e. the cells of the partition $P$.

\subsection{Meshing of dynamics: examples and counterexamples, in physics and 
philosophy}\label{21B}
I now consider the way in which the dynamics at the two levels can mesh with each other, or 
fail to do so. Physics provides precise and important examples of such `meshing' (e.g. in 
statistical mechanics), as well as examples where it fails.  I also relate meshing to the 
views of Fodor and Papineau on multiple realizability, and to the views of List on free 
will.

The framework, especially assumptions (i) and (iv) of Section \ref{21A}, yields natural 
definitions of:\\
\indent (a) how the micro-dynamics $T$ defines a macro-dynamics, and \\
\indent (b) whether these two dynamics mesh. 

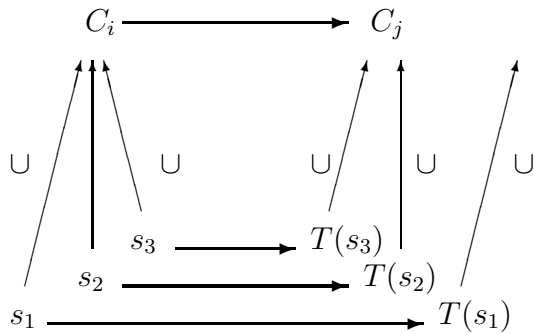
\begin{figure}[h] 
\setlength{\unitlength}{1mm}  
\centering      
\begin{picture}(60,50)   
\put(11,10){\vector(0,1){25}} 
\put(2,5){\vector(1,4){7.5}} 
\put(17.5,15){\vector(-1,4){5}} 
\put(52,10){\vector(0,1){25}}
\put(42.5,15){\vector(1,4){5}}
\put(60,5){\vector(1,4){7.5}}
\thicklines
\put(15,40){\vector(1,0){30}}  
\put(22,10){\vector(1,0){16}} 
\put(15,5){\vector(1,0){30}} 
\put(5,0){\vector(1,0){50}} 
\put(10,39){$C_i$} \put(48,39){$C_j$}
\put(0,0){$s_1$} \put(9,5){$s_2$} \put(16,10){$s_3$}
\put(0,20){$\cup$} \put(20,20){$\cup$} \put(40,20){$\cup$}
 \put(54,20){$\cup$} \put(67,20){$\cup$}
\put(57,0){$T(s_1)$} \put(47,5){$T(s_2)$} \put(40,10){$T(s_3)$}
\end{picture} 
\caption{Micro-determinism induces indeterministic macro-dynamics} 
\label{fig:noco}    
\end{figure}

As to (a), recall that any function $f:X \raw Y$ between any two sets $X$ and $Y$ defines a 
function ${\bar f}:{\cal P}(X) \raw {\cal P}(Y)$ between their power sets, by the obvious 
rule $A \subset X \mapsto f(A) : = \{ y \in Y \; | \; y = f(x) {\rm{\; for \; some \;}} x 
\in A \} \subset Y$. We apply this idea to $T: {\mathbb S} \raw {\mathbb S}$, and then 
restrict ${\bar T}$ to the partition $P = \{ C_i \} \subset {\cal P}({\mathbb S})$. In 
general, the image ${\bar T}(C_i)$ of a cell $C_i$ {\em is not} a subset of a single 
macro-state. That is: two distinct $s_1 \neq s_2 \in C_i$ are sent by $T$ to distinct 
macro-states. In other words: we have macro-{\em indeterminism}, despite $T$ giving 
micro-determinism. The micro and macro dynamics do {\em not} mesh. Cf Fig \ref{fig:noco}; 
where the union symbol, $\cup$, beside some of the upward lines indicates coarse-graining.

On the other hand, as to (b): if for every cell $C_i \in P$, its image  ${\bar T}(C_i)$ is 
a cell in $P$,  ${\bar T}(C_i) = C_j$ for some $j$, then the micro-dynamics $T$ {\em does} 
induce a deterministic macro-dynamics, and we will say that the micro- and macro-dynamics 
{\em mesh}. So in Fig \ref{fig:noco}  the upward $\cup$ arrow from $T(s_1)$ would not be 
`stray', but would point to $C_j$. In mathematical jargon: coarse-graining and 
time-evolution {\em commute}. Another jargon: coarse-graining is {\em equivariant} with 
respect to the group actions representing time-evolution; (i.e. for discrete time: actions 
of the group of integers  ${\mathbb Z}$, on ${\mathbb S}$ and on $P$).

Physics provides many important examples of meshing dynamics. The most obvious examples 
concern conserved quantities, especially in integrable systems. Recall from (iii) and (v) 
of Section \ref{21A} the idea of macro-quantities ${\mathbb Q}_{\rm{mac}} \subset {\mathbb 
Q}$, i.e. the idea that each cell $C_i$ is the intersection of level-surfaces, one for each 
quantity in ${\mathbb Q}_{\rm{mac}}$. If we instead define ${\mathbb Q}_{\rm{cons}} \subset 
{\mathbb Q}$ as those quantities whose values are constant under time-evolution $T$ (which 
could now even be indeterministic), then obviously, the intersection of the level-surfaces 
of these conserved quantities will be invariant under $T$. So there is meshing dynamics, 
though the macro-dynamics is trivial, i.e. every quantity we are concerned with is constant 
in value.   And one could go on to investigate the `integrable' systems for which ${\mathbb 
Q}_{\rm{cons}}$ is `rich enough', in the sense that these intersections are the minimal 
sets  invariant under time-evolution; cf. also Section \ref{21C}.

But there are important examples of meshing, when the system is not integrable, and even 
has only a few conserved quantities. Indeed, that is putting it mildly! Several of the most 
famous and fundamental equations of macroscopic physics (such as the Boltzmann, 
Navier-Stokes  and diffusion equations) are the meshing macro-dynamics induced by a 
micro-dynamics. Or rather: they are the meshing macro-dynamics once we make Section 
\ref{21A}'s framework more realistic by allowing that:\\
\indent  (a) the meshing may not last for all times;\\
\indent (b)  the meshing may apply, not for all micro-states $s$, but only for all except a 
`small' class;\\
\indent (c)  the coarse-graining may not be so simple as partitioning ${\mathbb S}$; and 
indeed \\
\indent (d) the definition of the micro-state space ${\mathbb S}$ may require approximation 
and-or idealization, especially by taking a limit of a parameter: in particular, by letting 
the number of microscopic constituents tend to infinity, while demanding of course that 
other quantities, such as mass and density, remain constant or scale appropriately.

This point, especially (d), returns us to Section \ref{redn}'s claim that  emergent (i.e. 
novel, robust) behaviour may be deduced from a theory of the microscopic details, often by 
taking a limit of some parameter. As I said there, my (2011a) gave four such examples. The 
equations just listed---Boltzmann etc.---provide others.\footnote{Uffink and Valente (2010)  
expound and assess the principal deduction of the Boltzmann equation, viz. Lanford's 
theorem. This example is especially topical in view of Villani's 2010 Fields medal (cf. 
Ambrosio (2011), Yau (2011)).} 

Having emphasized dynamical meshing, not least for its philosophical importance in 
reconciling emergence and reduction, I should add that on the other hand, {\em failure} of 
meshing---in mathematical jargon, the non-commutation of the diagram in Fig \ref{fig:noco}; 
in physical jargon, macro-indeterminism---need not pose any difficulties or `worries' for 
our understanding the situation. For there can be other factors that make the 
non-commutation, the macro-indeterminism, well understood, and even well controlled and 
natural. 

For a physicist, the obvious and striking example of this is the pilot-wave theory, in the 
foundations of quantum mechanics; as follows. The deterministic evolution $T$ on the 
pilot-wave micro-states $s$ (comprising e.g. positions ${\bf q}_i \in \mathR^3$ of 
corpuscles $i = 1,...,N$, as well as an orthodox quantum state $\psi$) induces an {\em 
indeterministic} evolution on $\psi$ by a precise version of  the textbooks' projection 
postulate rule that when $\psi$ divides into disjoint wave-packets, it is to be replaced by 
whichever of the (renormalized) packets has support containing the actual configuration 
$\langle {\bf q}_i \rangle \in \mathR^{3N}$.  Besides, a mathematically natural  
probability measure on the micro-states $s$ (dependent on the quantum state $\psi$: viz. $| 
\psi |^2$) makes the macro-indeterministic dynamics probabilistic, with the induced 
probabilities being the orthodox Born-rule probabilities (which are empirically correct for 
myriadly many kinds of experiments). So in this example, the macro-indeterminism is 
entirely understood, well controlled and natural; (cf. Bohm and Hiley (1992, especially 
Chapter 3), Holland (1993, especially Chapter 3)).

For a philosopher also, there is an obvious and striking example of macro-indeterminism, 
i.e. non-commutation of Fig \ref{fig:noco}. Namely: the venerable idea of free will; (ah, 
the joys of interdisciplinarity!). More precisely: a philosopher attracted by 
compatibilism---i.e. the view that determinism and free will are compatible---can take this 
sort of macro-indeterminism induced by a deterministic micro-dynamics to be exactly what 
free will involves. Or, still more precisely: what free will could be taken to involve in a 
world governed such a micro-dynamics. For a full defence of this version of compatibilism 
(including discussion of such background assumptions as non-reductive physicalism), I 
recommend List (2011).

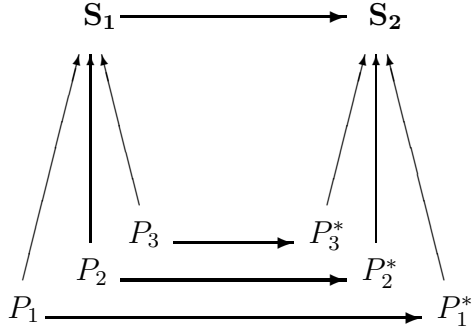
\begin{figure}[h] 
\setlength{\unitlength}{1mm}  
\centering      
\begin{picture}(60,50)   
\put(11,10){\vector(0,1){25}} 
\put(2,5){\vector(1,4){7.5}} 
\put(17.5,15){\vector(-1,4){5}} 
\put(49,10){\vector(0,1){25}}
\put(42.5,15){\vector(1,4){5}}
\put(58,5){\vector(-1,4){7.5}}
\thicklines

\put(15,40){\vector(1,0){30}}  
\put(22,10){\vector(1,0){16}} 
\put(15,5){\vector(1,0){30}} 
\put(5,0){\vector(1,0){50}} 
\put(10,39){$\bf{S}_1$} \put(48,39){$\bf{S}_2$}
\put(0,0){$P_1$} \put(9,5){$P_2$} \put(16,10){$P_3$}
\put(57,0){$P^*_1$} \put(47,5){$P^*_2$} \put(40,10){$P^*_3$}
\end{picture} 
\caption{Fodor's hope: meshing dynamics} 
\label{fig:Fodor}    
\end{figure}

On the other hand, I agree that in these less well-defined philosophical contexts about the 
relations between levels, failure of meshing can be `worrying'. One example of such a  
worry is Papineau's critique (2010, pp. 180-185) of Fodor's (1974) vision of special 
sciences as autonomous. Thus I agree that Papineau is right to press failure of meshing as 
a problem for Fodor. He argues that Fodor just assumes without justification that there 
will be meshing: Fodor's discussion appeals to a diagram like Fig \ref{fig:Fodor} which 
(with a trivial change from my notation) pictures the micro-laws as indeed preserving the 
macro-categories, i.e. as inducing by coarse-graining a well-defined 
dynamics.\footnote{Agreed, Fodor's and Papineau's  jargon is different from mine. Both 
authors talk about laws, and-or causal relations, within each of the two levels, not about 
dynamics; and about realization or supervenience as the relation between the laws and their 
instances, not about coarse-graining. But I submit these are only, or almost only, 
differences of jargon.

Another example of concern over failure of meshing in the context of the mental-physical 
relationship is Atmanspacher's discussion of the emergence of mental states from 
neurodynamics (2012, Sections 4.1, 6; cf. Atmanspacher and beim Graben 2007, Section 2).}

 Papineau goes on (p. 186f.) to argue---surely rightly---that in many cases, in everyday 
life and the special sciences, the meshing shown in Fig \ref{fig:Fodor} is secured by the 
macro-categories being defined by {\em selection processes}. 

I take up the topic of selection in Section \ref{five}'s discussion of Ellis. For the 
moment, I just note that Ellis also addresses the idea of meshing dynamics. He calls the 
commutation we see in Fig \ref{fig:Fodor} `coherent higher level dynamics', `effective same 
level action', and `the principle of equivalence of classes' (2008, p. 7-8, and Fig 3b, p. 
45). And, at least as I read him,  his typology of five kinds of top-down causation  
presupposes such meshing. So before turning to that typology, it is appropriate to discuss, 
albeit briefly, how one might secure such meshing; cf. Section \ref{21C}.

\subsection{Meshing secured by re-defining the macro-states}\label{21C}
One response to the failure of meshing is to re-define the macro-states, so as to secure 
it. I shall briefly consider this response from a general, and so mathematical, 
perspective. This will amount to considering how to get macro-determinism `by 
construction': or as one might gloss it less charitably, `by {\em fiat}'! So I should 
emphasise that in a real scientific context, one would usually invoke considerations about 
how to respond which are much more specific than the general ideas (like taking unions of 
the cells of the given partition) which I now mention. For example, one might invoke 
considerations like those in Section \ref{21B}: about conserved quantities, or about 
allowances (a)-(d), or about selection processes, as discussed by Papineau. 

I will make three comments. (1): The first is a ``false start''. (2): The second is a 
response to the false start; and (3): the third is a look at what one might call the 
`converse' to failure of meshing. All three will be recognizable as, in effect, some of the 
first steps in anyone's study of discrete-time dynamical systems. 

(1) {\em A false start}:--- Given a cell $C_i$ that ``gets broken up'' by the 
time-evolution $T$, there seem at first sight to be two possible tactics for changing the 
partition so as to secure meshing. But each runs into difficulties.

\indent (i): We might define a coarser notion of macro-state (a smaller partition with 
larger cells), by taking the union of the cells that contain images $T(s_1), T(s_2)$ of 
states $s_1, s_2$ in $C_i$ that get sent to different cells. That is: writing as usual 
$TC_i$ for the image-set, i.e. $TC_i := \{s' | s' = Ts , {\rm{\; some \;}} s \in C_i \; 
\}$, we might define
\be
[TC_i] : = \cup_{\; C_j \cap TC_i \; \neq \; \emptyset\;} \; C_j \; .
\ee
However, the sets $TC_i$, as $C_i$ runs through $P$, are not a partition of $\mathbb S$, 
since different $TC_i$ can overlap. So to get a partition with a meshing dynamics, we have 
to define a {\em chain} of overlapping sets $TC_i$, i.e. a sequence $\langle TC_0, 
TC_1,..., TC_N \rangle$, with $TC_0 \cap TC_1 \neq \emptyset, TC_1 \cap TC_2 \neq 
\emptyset,$ etc; and then define the partition whose cells are maximal chains of image-sets 
$TC_i$. This partition has, by construction, a meshing dynamics. But it is liable to be 
``uninformative'': that is, the unions of maximal chains of image-sets $TC_i$ are liable to 
be large.

\indent (ii):  We might define a finer notion of macro-state (a larger partition with 
smaller cells), by decomposing the cell $C_i$ that gets broken up by  $T$ into subsets, 
according to which cells its elements $s \in C_i$ get sent to. That is: we might define, 
for each $j$ in the index set of the given partition $P$
\be
(TC_i)_j : = C_i \cap T^{-1}(C_j) \; ;
\ee
and then consider the decomposition of $C_i$ into its subsets $(TC_i)_j$.  Then the sets 
$(TC_i)_j$, as $i, j$ run through the index set of $P$, obviously form a partition of 
$\mathbb S$ (perhaps with, harmlessly, various ``copies'' of the empty set, i.e. cases 
where $(TC_i)_j = \emptyset$). And since this partition has smaller cells than $P$ did, 
this tactic is not in danger of being uninformative in the way that the first tactic, in 
(i), was.  But now the trouble is that this partition need not have a meshing dynamics. For 
though indeed $T$ maps all of $(TC_i)_j$ into $C_j$, $T$ need not map all of  $(TC_i)_j$ 
into some cell of the partition we have just defined, i.e. into some  $(TC_k)_l$. 

(2) {\em A response}:--- Rather than starting from a given partition $P$ with a non-meshing 
dynamics, and asking how to modify it so as to get meshing, it is easier to start with just 
the micro-dynamics and consider defining {\em ab initio} a partition with a meshing 
dynamics. Indeed, it is easy to address the stronger question of defining cells each of 
which is invariant under the dynamics, i.e. mapped into themselves by $T$. 

Thus for any state $s \in {\mathbb S}$, the set
\be
[Ts] : = \{s' | {\rm{\; there \; is \; a \; chain \;}} s_0 = s, s_1 = Ts, ..., s_n = 
Ts_{n-1}, ..., s_N = s' \; \}
\label{Ts}
\ee
is obviously the smallest set invariant under $T$ that contains $s$. This statement also 
holds true if $T$ is not a function, but one-many, i.e. the time-evolution is 
indeterministic (so that many chains could start at $s$). If $T$ {\em is} a function, then 
$[Ts]$ is either a cycle, i.e. $s \mapsto s_1 \mapsto s_2 \mapsto ... \mapsto s$, or is an 
$\omega$-sequence.

By containing only ``descendants'' of $s$, eq. \ref{Ts}'s definition of $[Ts]$ obviously 
favours the ``initial time''. If $T$ is a function, we can avoid this favouritism by 
instead defining 
\begin{eqnarray}
[[Ts]] : = \{s' | {\rm{\; either \; there \; is \; a \; chain \;}} s_0 = s, s_1 = Ts, ..., 
s_n = Ts_{n-1}, ..., s_N = s' \; , \nonumber \\
{\rm{\; or \; there \; is \; a \; chain \;}} s_0 = s', s_1 = Ts', ..., s_n = Ts_{n-1}, ..., 
s_N = s \}
\label{Ts1}
\end{eqnarray}
which is obviously the smallest set invariant under $T$ that contains both $s$ and any of 
its ``ancestors''. If T is not a function, then we need to allow for chains from ancestors 
of $s$ that do not pass through $s$. Then
\be
[[[Ts]]] := [[Ts]] \cup \cup_{s' \in [[Ts]]} [Ts']
\ee
is by construction the smallest set invariant under $T$ that contains both $s$ and any of 
its ``ancestors''.

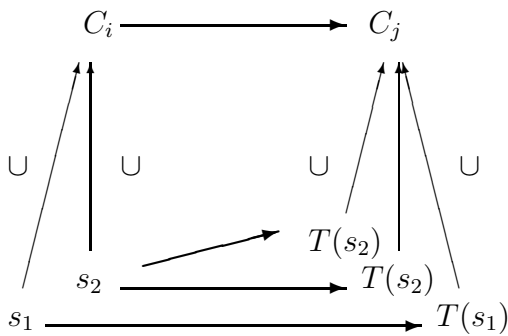
\begin{figure}[h] 
\setlength{\unitlength}{1mm}  
\centering      
\begin{picture}(60,50)   
\put(11,10){\vector(0,1){25}} 
\put(2,5){\vector(1,4){7.5}} 
\put(52,10){\vector(0,1){25}}
\put(45,15){\vector(1,4){5}}
\put(60,5){\vector(-1,4){7.5}}
\thicklines
\put(15,40){\vector(1,0){30}}  
\put(18,08){\vector(4,1){18}} 
\put(15,5){\vector(1,0){30}} 
\put(5,0){\vector(1,0){50}} 
\put(10,39){$C_i$} \put(48,39){$C_j$}
\put(0,0){$s_1$} \put(9,5){$s_2$} 
\put(0,20){$\cup$} \put(15,20){$\cup$} \put(40,20){$\cup$}
 \put(60,20){$\cup$}
\put(57,0){$T(s_1)$} \put(47,5){$T(s_2)$} \put(40,10){$T(s_2)$}
\end{picture} 
\caption{Micro-indeterminism induces deterministic macro-dynamics} 
\label{fig:conversenoco}    
\end{figure}

(3) {\em The converse scenario}:--- So much by way of discussing the failure of meshing, 
i.e. the scenario in Fig \ref{fig:noco}, and how one might respond to it by re-defining the 
macro-states. That scenario also prompts one to consider the converse scenario: that is, 
micro-indeterminism inducing, by coarse-graining, macro-determinism, as in Fig 
\ref{fig:conversenoco}. 

Indeed, this scenario can happen in a tightly defined theoretical context, in a 
scientifically important way. This means in particular: the macro-determinism is not ``won 
on the cheap'', by having very large cells (in other words: by using a logically weak 
taxonomy of micro-states). Brownian motion provides an example. Think of the Langevin 
equation's probabilistic description of Brownian motion. Different realizations of the 
noise (i.e. different, unknown, trajectories of the atoms bombarding the large Brownian 
particle) give the particle different spatial trajectories: micro-indeterminism. But 
averaging over the realizations give a deterministic evolution of the probability density 
for the particle's position; (this evolution is given by a Fokker-Planck 
equation).\footnote{Probability theory also provides similar examples where time-evolution 
is downplayed, but which retain the key idea of combining variety of micro-states with 
macroscopic uniformity. One main example is the method of arbitrary functions; where the 
probability  of a macro-state is approximately the same for any of a wide class of 
probability density functions $f$ on the micro-states $s \in {\mathbb S}$, essentially 
because the partition $P$ defining the macro-states is very intricate. Butterfield (2011a, 
Section 4) gives more details, including a discussion of emergence.}

\section{Ellis'  types of top-down causation}\label{five}
I turn to using the framework of Section \ref{mesh} to describe, in part, the first four of 
Ellis' five types of top-down causation (Ellis 2008, 2012). I think these descriptions are 
worth formulating, because they are precise. But I say `in part', because this precision 
comes with a price-tag; indeed two price-tags.

 First, I will face several choices about how best to formalize Ellis' ideas. For clarity 
and brevity, I will make simple choices: even to the extent of {\em not} representing 
micro-states and micro-dynamics, and so also not representing top-down causation. But I 
shall suggest my choices are innocuous: one can see how one could elaborate the 
descriptions so as to represent micro-states, top-down causation etc. 

But second, and perhaps more important, I will also simplify by setting aside non-formal, 
indeed philosophical, questions about what, beyond ideas like functional dependence and 
coarse-graining, is needed for causation, in particular top-down causation. (Recall my 
concessions (ii) and (iii) at the end of Section \ref{intro}.)  For example, I set aside: 
the distinction, made by Ellis (2008, p. 6, p.8 et seq.) and Auletta et al. (2007), between  
causal effectiveness and causal power; and their taking top-down causation to require the 
macro-level to have causal power over the micro-level. And apart from the views of Ellis 
and kindred spirits like Jaeger and Calkins: I will not try to connect my descriptions of 
Ellis' third and fourth types, which concern adaptation and evolution and so biology, to 
biological details, or to this issue's other relevant authors (Love (2011), Noble (2011), 
Okasha (2011)).

These simplifications are evident already in my proposed description of Ellis' first type, 
which he calls {\em algorithmic top-down causation}. Ellis writes (2008, p. 8): 
`algorithmic top-down causation occurs when high-level variables have causal power over 
lower level dynamics through system structuring, so that {\em the outcome depends uniquely 
on the higher level structural, boundary and initial conditions}'. In line with my 
simplifications, I will take this quotation to mean {\em just} meshing dynamics, or 
commutation, as in Section \ref{21B}: the `Fodor's hope' of Fig \ref{fig:Fodor}. For 
meshing dynamics matches the quotation's idea of a higher level outcome being uniquely 
determined by higher level facts. But that this is indeed a simplification is clear from 
 Ellis also calling meshing dynamics `coherent higher level action emerging from lower 
level dynamics' and `the principle of equivalence of classes' (ibid.), and his taking it as 
a presupposition of his typology of top-down causation.

 I will devote a Subsection (\ref{22B} to \ref{22D}) to describing each of Ellis' second, 
third and fourth types. For clarity, I will begin each Subsection with Ellis' own 
description (in his 2008) of the type. For brevity, I omit his fifth type (called 
`intelligent top-down causation') which involves the use of
symbolic representation to investigate the outcome of goal choices. But by the end of my 
description of his fourth type (Section \ref{22D}), it will be clear how the fifth type 
might be partially described with my framework.

\subsection{Top-down causation via non-adaptive information control}\label{22B}
Ellis writes (2008, p. 12): `in non-adaptive information control, higher level entities 
influence lower level entities so as to attain specific fixed goals through the existence 
of feedback control loops, whereby information  on the difference between the system's 
actual state and desired state is used to lessen this discrepancy ... unlike [algorithmic 
top-down causation], {\em the outcome is not determined by the boundary or initial 
conditions; rather it is determined by the goals}'.

I will represent this notion of control in my framework, using two simple ideas (which will 
be useful later): which I will call {\em culling} and {\em conforming}. In both culling and 
conforming, some macro-state $C^*$ (cell $C^*$ of the partition $P$) is designated as 
`desired'. The difference will be that in culling, one simply sends to some ``rubbish'' or 
``dead'' state, 0 say, all states other than the desired one $C^*$. So this will be 
modelled with a characteristic function $\chi_{C^*}$ (though with the ``yes'' or 
``winning'' value being $C^*$ itself, rather than 1). In conforming, on the other hand, 
states other than the desired one are sent to the desired one: so this will be modelled by 
a constant map with value $C^*$. The details are as follows, with due precision about the 
difference between maps at the macro- and micro-levels.

{\em Culling}: Writing 0 for the rubbish state at the macro- or micro-level, and adjoining 
the rubbish macro-state to the partition $P$ so as to define a partition $P^0 := P \cup \{ 
0 \}$, we have:---\\
\indent (a) at the macro-level: a characteristic map $\chi_{C^*}: P \raw P^0$; with 
$\chi_{C^*}(C_i) := 0$, unless $C_i = C^*$ in which case $\chi_{C^*}(C_i) := C^* \equiv 
C_i$.\\
\indent (b) at the micro-level: a characteristic map $\chi_{C^*}: {\mathbb S} \raw \{ 
{\mathbb S}, 0 \}$, or if one prefers $\chi_{C^*}: {\mathbb S} \raw \{ C^*, 0 \}$, with a 
similar ``culling'' definition.

{\em Conforming}: We have:---\\
\indent (a) at the macro-level: a constant  map $\kappa_{C^*}: P \raw P$ sending all cells 
in $P$ to $C^*$: $\kappa_{C^*}(C_i) := C^*$, for all $C_i$.\\
\indent (b) at the micro-level: {\em any} micro-dynamics given by a function (or indeed 
relation, i.e. multi-valued function) $T$ on $\mathbb S$ that sends all of $\mathbb S$ into 
$C^*$. That is:  any $T$ such that $T({\mathbb S}) \subset C^*$ will induce $\kappa_{C^*}: 
P \raw P$ as its meshing macro-dynamics.

Various combinations and liberalisations of these two ideas are possible. The simplest 
combination is to conform and then cull: $\chi_{C^*} \circ  \kappa_{C^*}$. Then no $s \in 
{\mathbb S}$ ends up as ``rubbish''. Instead, all $s \in {\mathbb S}$ get sent to a 
micro-state in the desired macro-state $C^*$. This is a perfect control, with the system 
ending in the desired macro-state, whatever its initial $s \in {\mathbb S}$.

\subsection{Top-down causation via adaptive selection}\label{22C}
Ellis writes (2008, p. 14-15): `Adaptive processes take place when many entities interact 
... for example individuals in a population, and {\em variation takes place in the
properties of these entities, followed by selection of preferred entities that are better
suited to their environment or context}. Higher level environments provide
niches that are either favorable or unfavorable to particular kinds of lower level
entities; those variations that are better suited to the niche are preserved and the others
decay away. Criteria of suitability in terms of fitting the niche can be thought of as
fitness criteria guiding adaptive selection. On this basis a {\em selection agent} or {\em 
selector} accepts one of the states and rejects the rest; this
selected state is then the current system state that forms the starting basis for the next
round of selection ... Thus this is top-down causation from the context to the system. An 
equivalence class
of lower level variables will be favored by a particular niche structure in association
with a specific fitness criteria. Unlike feedback control, this process does not attain 
preselected
internal goals by a specific set of mechanisms or systems; rather it creates
systems that favor the meta-goals embodied in the fitness criteria.'

In representing these ideas, one of course faces several choices about how much detail to 
represent explicitly. For example, should one represent explicitly:\\
\indent (i) the environment/context (even niches?), or just the adapting entities; \\
\indent (ii) as regards these entities: individuals (and their variation?), or just the 
population; \\
\indent (iii) as regards the environment and the entities: micro-states, or just 
macro-states;  \\
\indent (iv) selection as a process with several possible outcomes, or just two (survival 
or death!), as in Section \ref{22B}'s notion of culling; \\
\indent (v) several rounds of selection, or just one;\\
\indent (vi) adaptation using Section \ref{22B}'s notion of conforming;  \\
\indent (vii) adaptation {\em within} a round of selection, e.g. in the lifetime of an 
individual, or just over many rounds? 

In answering these questions, I propose to keep things pretty simple. In terms of this 
list, I will represent explicitly:\\
\indent \indent (i) the environment;\\
\indent \indent  (ii) individuals and their variation\\
\indent \indent  (v) several rounds of selection. \\
But I will {\em not} represent: \\
\indent \indent  (iii) micro-states; \\
\indent \indent  (iv) outcomes of selection other than survival or death, as in culling;\\
\indent \indent  (vi) and (vii): adaptation, in terms of conforming, in a single round or 
over many rounds. \\
But I admit that these are just {\em choices} of what to represent: no doubt, several other 
possible choices are equally (or more) useful. 

As to (i), the environment: I will be simplistic. I take the environment to be unchanging, 
so that there is no co-evolution. There is an environment state-space ${\mathbb S}^e$ on 
which there is a partition $P^e = \{ C^e_k \}$. Niches will be represented only implicitly, 
viz. by the way in which the macro-states of the individuals (and so of the population) 
that are selected for depend on the macro-state $C^e_k$ of the environment. 

As to (ii), the individuals and their variation: again, I will be simplistic. I assume that 
in each round of selection, there are $N$ individuals, each with an individual state-space 
$\mathbb S$ on which there is a partition $P = \{ C_i \}$. So the population of $N$ 
individuals has a Cartesian product state-space ${\mathbb S}^N$ (neglecting the tensor 
products of quantum theory!), with the product partition $P^N$ whose cells are given by 
$N$-tuples $\langle C_{i_1}, C_{i_2}, ..., C_{i_N} \rangle$. So variation consists in not 
all individuals being in one cell: i.e. the population macro-state is not an $N$-tuple with 
all components equal to some single $C_i$. (Here one could adopt the occupation number 
formalism from statistical physics.) 

Finally, as to (v), several rounds of selection: again, I will be simplistic. Not only will 
I take selection to be essentially  culling as in Section \ref{22B}, albeit with a 
designated/desired macro-state $C^*$ that is a function of the environment's macro-state, 
i.e. $C^* = C^*(C^e_k)$. Also, I will assume that after culling:\\
\indent (a) the $M$ ($M \leq N$) surviving individuals simply persist in the desired 
macro-state $C^*$ that they (are so lucky to!) have been in; (or if you prefer: each is 
replaced by an offspring in the macro-state $C^*$); and \\
\indent (b) $N - M$ new individuals spring up (as from dragons' teeth!), in some randomly 
selected combination of macro-states, to proceed to the next culling, along with the $M$ 
individuals in $C^*$ who have survived from the last round. 

Now it is obvious how this toy-model achieves adaptation. Since I have assumed that the 
environment is unchanging, i.e. always in a certain macro-state $C^e_k$, the culling always 
favours the same macro-state, $C^* = C^*(C^e_k)$, of individuals. Therefore, over 
sufficiently many rounds of selection, the random production, at the start of each round, 
of unfit variations, i.e. macro-states $C_i \neq C^*(C^e_k)$, decays away. That is: over 
time, the population (and the individuals) achieves adaptation in the sense of Section 
\ref{22B}'s notion of conforming: i.e. a constant map taking the desired macro-state $C^*$  
as its value.

This scenario can be summed up in terms of functions on macro-states. First, we adapt to 
the partition $P^N$ the notation we used in Section \ref{22B} for culling. That is:\\
\indent (a) we adjoin the rubbish macro-state 0 to the partition $P^N$ so as to define a 
partition $P^{(N,0)} := P^N \cup \{ 0 \}$; and \\
\indent (b) we adopt the obvious component-wise definition of the characteristic function 
$\chi_{C^*} \equiv \chi_{C^*(C^e_k)} : P^N \raw P^{(N,0)}$, defined on the partition $P^N$ 
and with codomain $P^{(N,0)}$.

We also need to represent the birth, after each round of culling, of the new individuals. 
We do this by postulating that after each round, we apply a function $\beta: P^{(N,0)} \raw 
P^N$, which is defined (i) to keep constant any component equal to $C^*$, and (ii) to 
replace any component equal to 0 by some ``living'', non-rubbish macro-state $C_i \in P$ 
(as hinted by $\beta$'s codomain being just $P^N$). But I shall not formalize the idea that 
the new individuals' macro-states, given by (ii) of $\beta$, are randomly chosen: I shall 
simply imagine that for each round of culling, the function $\beta$ is in general 
different; so that we postulate a sequence of functions, $\beta_1, \beta_2, \beta_3, ...$, 
each subject to the requirements (i) and (ii).

Thus each round of selection and birth is an application of the characteristic culling 
function $\chi_{C^*(C^e_k)}$, followed by an application of one of the $\beta$ functions, 
representing birth of new individuals. So the (macro)-histories of all the individuals, and 
of the population, that are possible, for a given environment macro-state $C^e_k$ and given 
random functions $\beta_n$, are encoded in the sequence of functions:
\be
P^N \stackrel{\chi_{C^*(C^e_k)}}{\longrightarrow} P^{(N,0)} 
\stackrel{\beta_1}{\longrightarrow} P^N 
\stackrel{\chi_{C^*(C^e_k)}}{\longrightarrow} P^{(N,0)} \stackrel{\beta_2}{\longrightarrow} 
P^N
\stackrel{\chi_{C^*(C^e_k)}}{\longrightarrow} P^{(N,0)} \stackrel{\beta_3}{\longrightarrow} 
\; \cdots \; \stackrel{\beta_p}{\longrightarrow} P^N \; .
\ee 
These functions are defined so that a generic initial population macro-state is mapped 
eventually to a state where all individuals are in $C^*$: adaptation! Thus suppose that in 
the first generation, the first, third and $N$th individuals happen to be in the desired 
macro-state $C^*$, while the second, fourth (and no doubt other!) individuals are not and 
so get culled. And suppose that after many, say $p$, generations, the functions $\beta_n$ 
have been ``random enough'' to have thrown up at some time (i.e. for some $n$) the desired 
macro-state $C^*$ for every component.  This would give a (macro)-history as follows:
\begin{eqnarray}
\langle C_{i_1}, C_{i_2}, C_{i_3}, C_{i_4}, ..., C_{i_N} \rangle \equiv 
\langle C^*, C_{i_2}, C^*, C_{i_4}, ..., C^* \rangle 
\stackrel{\chi_{C^*(C^e_k)}}{\longmapsto}
\nonumber \\
\langle  C^*, 0,  C^*, 0,...,  C^* \rangle 
\stackrel{\beta_1}{\longmapsto}
\langle C^*, C'_{i_2}, C^*, C'_{i_4}...,  C^* \rangle 
\stackrel{\chi_{C^*(C^e_k)}}{\longmapsto} 
\nonumber \\
\cdots 
\stackrel{\beta_p}{\longmapsto}
\langle C^*, C^*, C^*,..., C^* \rangle .
\end{eqnarray}

To sum up this Section: here is a toy-model of adaptive selection. Agreed, it is very 
simple. Indeed, it does not represent micro-states; so that in terms of my framework, it 
cannot represent top-down causation. But I  submit that if we elaborated the model so as to 
include micro-states, we would be in a situation like that in Section \ref{22B}. We would 
face issues about whether the model's macro-dynamics meshes with its micro-dynamics. For 
example,  non-meshing would  be threatened by indeterminism, due to each individual, and so 
the population as a whole, being an open system. But we could write down a meshing dynamics 
(perhaps, for realism, availing ourselves of allowances like (a) to (d) in Section 
\ref{21B}); and thus get a formal description of Ellis' third type of top-down 
causation---at least in the sense of causation as functional dependence.

\subsection{Top-down causation via adaptive information control}\label{22D}
Ellis writes (2008, p. 18): `Adaptive information control takes place when there is 
adaptive selection of goals in a
feedback control system, thus combining both feedback control [this paper's Section 
\ref{22B}] and
adaptive selection [this paper's Section \ref{22C}]. The goals of the feedback control 
system are
irreducible higher level variables determining the outcome, but are not fixed as in the
case of non-adaptive feedback control; they can be adaptively changed in response to
experience and information received. The overall process is guided by fitness criteria for
selection of goals, and is a form of adaptive selection in that goal selection relates to
future rather then present use of the feedback system. This allows great flexibility of
response to different environments, indeed in conjunction with memory it enables
learning and anticipation ... and underlies effective purposeful
action as it enables the organism to adapt its behaviour in response to the environment in
the light of past experience, and hence to build up complex levels of behaviour.'

In representing these ideas, one again needs to make choices, and exercise judgment, about 
which details to be explicit about. I will again be simplistic. I will also build on 
Section \ref{22C}'s choices, and its ensuing notations. This means that, although I will 
represent the idea that the goal is not fixed but depends on history, I will {\em not} 
capture one prominent feature of control or feedback: viz. the system's time-evolution 
being guided towards the goal, by for example the system calculating the difference between 
its present state and the goal-state, and then engineering its change in the next time-step 
so as to reduce this difference.

Instead, I will postulate, like I did in Section \ref{22C}, that a system evolves randomly, 
until it happens to hit its goal---after which it stays in that state. Agreed, that does 
not merit the names `control', `feedback' or `learning'. But it will be clear that, at the 
cost of more definitions and notation, my framework could equally well describe these 
notions, namely in terms of time-evolutions using such difference-reducing rules. Thus I 
submit that what follows, though simple, is enough for my purpose, viz. to show one could 
modify a model of evolution such as Section \ref{22C}'s so as to represent Ellis' `adaptive 
information control'. 

Recall that in Section \ref{22C}, the goal $C^*$ was a function of just the (time-constant) 
environment macro-state $C^e_k$. We had $C^* = C^*(C^e_k)$; and throughout the process of 
evolution (and adaptation), the same $C^*$ acted as the goal (the attractor macro-state for 
each individual). I now modify this so as to make each individual's goal a function of its 
history, and also the environment's history; (recall Ellis' mention of memory and past 
experience). 

So let us imagine $N$ persisting individuals (despite Section \ref{22C}'s talk of offspring 
and births), labelled by $j = 1,2,..., N$. Time is discrete, with the generic time-point 
labelled $n$. So the environment passes through a sequence of macro-states
\be
C^e_{k_1} \longmapsto C^e_{k_2} \longmapsto   C^e_{k_3} \;\; \cdots \;\; \longmapsto 
C^e_{k_n} \longmapsto \;\; \cdots \; ;
\ee 
and individual $j$ passes through a sequence of macro-states
\be
C'_{i_j} \longmapsto  C''_{i_j} \longmapsto  C'''_{i_j} 
\;\; \cdots \;\; \longmapsto C^{(n)}_{i_j} \longmapsto \;\; \cdots \; .
\ee 
We postulate that the goal of each individual $j$ at each time-point $n$ is given by a 
goal-function $C^*_{n;j}$ which takes as its argument a prior (macro)-history of both $j$ 
and of the environment (but for simplicity: not other individuals!), and as its value an 
individual's macro-state, i.e. an element $C_i$ of the partition $P$ of the individual 
state-space. 

So let us write $H_{n;j}$ for a (macro)-history of both $j$ and of the environment up to 
the time-point $n$, i.e. $H_{n;j}$ is a $2n$-tuple $\langle C'_{i_j}, C''_{i_j}, ..., 
C^{(n)}_{i_j} ; C^e_{k_1}, C^e_{k_2}, ..., C^e_{k_n} \rangle$, and let us write ${\cal 
H}_{n;j}$ for the class of these $2n$-tuples. Summing up: we have the goal-function 
\be
C^*_{n;j}: {\cal H}_{n;j} \ni H_{n;j} = \langle C'_{i_j}, C''_{i_j}, ..., C^{(n)}_{i_j} ; 
C^e_{k_1}, C^e_{k_2}, ..., C^e_{k_n} \rangle \longmapsto C^*_{n;j}(H_{n;j}) \in P \; .
\ee  

 We now adjoin these definitions to Section \ref{22C}'s scenario. There, each time-step 
(`round')  involved a culling and rebirth for those individuals that had not attained their 
goal-state, while those in the goal-state $C^*$ simply persisted in it. Now, using our 
present metaphor of $N$ persisting individuals with an ever-lengthening history, each round 
must (i) keep those individuals that have attained their goal-state in that state, and (ii) 
assign to any other individual some random macro-state as its next state. 

The combination of (i) and (ii) means that, provided (ii)'s assignments of macro-states are 
sufficiently random that for each individual they sometimes assign its present goal-state, 
then in the long run, all the individuals attain---and stay in---their goals. In short: 
again, we have adaptation.

As regards spelling out the formal details of (i) and (ii): I will skip the details about 
(ii), which are a straightforward adaptation of the $\beta$ (``birth'') functions of 
Section \ref{22C}. As to (i), there are two features we need to require, the first being 
more important.\\
\indent (a): If up to  time-point $n + 1$, the individual $j$ and environment has 
experienced the joint history $H_{n;j}$, so that $j$'s goal is then $C^*_{n;j}(H_{n;j})$, 
and if $j$ happens to be in the state  $C^*_{n;j}(H_{n;j})$, then we require that $j$ will 
forever remain in $C^*_{n;j}(H_{n;j})$. That is: we require that once any individual enters 
its goal, it stays forever.\\
\indent (b): It is natural to have goal-functions ``stay consistent'', i.e. go on endorsing 
any goal that is attained. That is: it is natural to require that for any $j$ and $n$, and 
joint history $H_{n;j}$, with initial segments $H_{m;j}$ ($m \leq	n$): if the $m$th 
component of (the individual-history first-half of) $H_{n;j}$ is $C^*_{m;j}(H_{m;j})$, then 
all the later initial segments of $H_{n;j}$, i.e. $H_{p;j}$ with $p > m$, yield 
endorsements of this goal, i.e.  $C^*_{p;j}(H_{p;j}) = C^*_{m;j}(H_{m;j})$. (Due to (a), 
the system will in fact stay in $C^*_{m;j}(H_{m;j})$ at all times $p > m$).\\
\indent To sum up: these two points encode the idea that a goal-state is an attractor. \\ 
\\

\noindent {\em Acknowledgements}:-- I thank: the John Templeton Foundation and George Ellis 
for the invitation; Harald Atmanspacher, Robert Bishop, Adam Caulton, and especially 
Christian List and Elliott Sober for comments; the symposium participants for discussion; 
and above all, George Ellis for the inspiration of his work, and for very helpful detailed 
comments on an earlier version.

\section{References}

Amann, A. (1993), `The {\em Gestalt} problem in quantum theory: generation of molecular 
shape by the environment', {\em Synthese} {\bf 97},pp. 125-256.

Ambrosio, L. (2011), `The work of Cedric Villani', {\em Notices of the American 
Mathematical Society} {\bf 58}, pp. 464-468.

Atmanspacher, H. (2012), `Identifying mental states from neural states under mental 
constraints', {\em Interface Focus} (Royal Society, London), this issue. 

Atmanspacher, H. and beim Graben, P. (2007), `Contextual emergence of mental states from 
neurodynamics', {\em Chaos and Complexity Letters} {\bf 2}, pp. 151-168.

Auletta, G., Ellis, G. and Jaeger, L. (2008), `Top-Down Causation by Information Control: 
From a Philosophical Problem to a Scientific Research Program', {\em  Journal of the Royal 
Society Interface} {\bf 5}, pp. 1159-1172. arXiv:0710.4235.

Bedau, M. (2003), `Downward causation and autonomy in weak emergence', {\em Principia 
Revista Internacional de Epistemologica} {\bf 6}; reprinted in Bedau and Humphreys (2008), 
page reference to reprint. 

Bedau, M. and Humphreys, P. (eds.) (2008), {\em Emergence: contemporary readings in 
philosophy and science}, MIT Press: Bradford Books.

Bishop, R. and Atmanspacher, H. (2006), `Contextual emergence in the description of 
properties', {\em Foundations of physics} {\bf 36}, pp. 1753-1777.

Bohm, D. and Hiley, B. (1992), {\em The Undivided Universe}, Routledge. 

Butterfield, J. (2011), `Emergence, reduction and supervenience: a varied landscape', {\em  
Foundations of Physics} {\bf 41(6)}, pp. 920-960; doi:10.1007/s10701-011-9549-0; available 
at: http://philsci-archive.pitt.edu/archive/00005549/

Butterfield, J. (2011a), `Less is Different: emergence and reduction reconciled', {\em 
Foundations of Physics} {\bf 41(6)}, pp. 1065-1135;; doi:10.1007/s10701-010-9516-1;
available at: http://philsci-archive.pitt.edu/archive/00008355/

Dennett, D.  (1991), `Real Patterns', {\em The Journal of Philosophy} {\bf 87}, pp. 27-51;  
reprinted in Bedau and Humphreys (2008), page reference to reprint. 

Ellis, G. (2008), `On the nature of causation in complex systems', 
{\em Transactions of the Royal Society of South Africa} {\bf 63} pp. 69-84; available at:
http://www.mth.uct.ac.za/~ellis/Top-down20Ellis.pdf; page references to latter version.

Ellis, G. (2011), `Top-down causation, emergence and transcendence: some comments on 
mechanisms', this issue of {\em Interface Focus}.

Fodor, J. (1974), `Special Sciences (Or: the disunity of science as a working hypothesis), 
{\em Synthese} {\bf 28}, pp. 97-115; reprinted in Bedau and Humphreys (2008).

Hempel, C. (1966), {\em Philosophy of Natural Science}, Prentice-Hall.

Hendry, R. (2010), `Ontological reduction and molecular structure', {\em Studies in the 
History and Philosophy of Modern Physics} {\bf 41}, pp. 183-191.

Holland, P. (1993), {\em The Quantum Theory of Motion}, Cambridge University Press.

Humphreys, P. (1997),`How properties emerge', {\em Philosophy of Science} {\bf 64}, pp. 
1-17; reprinted in Bedau and Humphreys (2008).

Kim, J. (1999), `Making sense of emergence', {\em Philosophical Studies} {\bf 95}, pp. 
3-36; reprinted in Bedau and Humphreys (2008), page reference to reprint.

Lewis, D. (1973), `Causation', {\em Journal of Philosophy} {\bf 70}, pp. 556-567.

List, C. (2011), `Free Will, determinism, and the possibility to do otherwise', available 
at: http://personal.lse.ac.uk/list/PDF-files/FreeWill.pdf

List, C. and Menzies, P. (2009), `Non-Reductive physicalism and the limits of the exclusion 
principle', {\em Journal of Philosophy} {\bf 106}, pp. 475-502.

Love, A. (2011), `A pluralist perspective on top-down causation', this issue of {\em 
Interface Focus}. 

Menzies, P. and List, C. (2010), `The causal autonomy of the special sciences', in C. 
Macdonald and G. Macdonald (eds), {\em Emergence in Mind}, Oxford University Press: pp. 
139-168.

Nagel, E. (1961), {\em The Structure of Science: Problems in the Logic of Scientific 
Explanation}, Harcourt.

Noble, D. (2011), `A theory of biological relativity: no privileged level of causation', 
this issue of {\em Interface Focus}.  

O'Connor, T. (2011), `A causally unified world in which fundamental physics is not causally 
complete', this issue of {\em Interface Focus}.

Okasha, S. (2012), `Emergence, hierarchy and top-down causation in evolutionary biology', 
this issue.

Papineau, D. (2010), `Can Any Sciences be Special?', in C. Macdonald and G. Macdonald 
(eds), {\em Emergence in Mind}, Oxford University Press; pp. 179-197; available at:\\
  http://www.kcl.ac.uk/content/1/c6/04/17/78/SpecialSciencesinMacdonalds.pdf

Primas, H. (1981), {\em Chemistry, Quantum Mechanics and Reductionism}, Berlin: Springer 
(second edition 1983). 

Putnam, H. (1975), `Philosophy and our mental life', in his collection {\em Mind, Language 
and Reality}, Cambridge University Press, pp. 291-303.

Scerri, E. (2011), `Top-down causation regarding the chemistry-physics interface', this 
issue of {\em Interface Focus}. 

Schaffner, K. (2011), `Ernest Nagel and reduction', forthcoming in {\em The Journal of 
Philosophy}.

Shapiro, L. (2000), `Multiple realizations', {\em The Journal of Philosophy} {\bf 97}, pp. 
635-654.

Shapiro, L., and Sober, E. (2007), `Epiphenomenalism --- the Do's and the Don'ts', in
P. Machamer and G. Wolters (eds.), {\em Thinking about Causes}, University of Pittsburgh 
Press, pp. 235-264.

Sober, E. (1999), `The multiple realizability argument  against reductionism',  {\em 
Philosophy of Science} {\bf 66}, pp. 542-564.

Sober, E. (1999a), `Physicalism from a probabilistic point of view',  {\em Philosophical 
Studies} {\bf 95}, pp. 135-174.

Uffink, J. and Valente, G. (2010), `Time's arrow and Lanford's theorem', {\em Seminaire 
Poincare: XV: Le temps}, pp. 141-173.

Yablo, S. (1992), `Mental causation', {\em The Philosophical Review} {\bf 101}, pp. 
245-280.

Yau, H-T. (2011), `The work of Cedric Villani', forthcoming in {\em Proceedings of the 2010 
International Congress of Mathematicians}. 

\end{document}